\documentclass[
 aps,
 prl,
 amsmath,amssymb,amsfonts,
 reprint,
 10pt,
 superscriptaddress,
 floatfix,
 longbibliography,
 noeprint
]{revtex4-2}

\usepackage{physics}

\usepackage[english]{babel}
\usepackage[utf8]{inputenc}
\usepackage[T1]{fontenc}
\usepackage{mathrsfs}
\usepackage{bbm}
\usepackage{bm}
\usepackage{enumerate}
\usepackage{mathtools}
\usepackage{graphicx}
\usepackage[dvipsnames]{xcolor}
\usepackage{float}
\usepackage{subdepth}
\usepackage{booktabs}
\usepackage{multirow}
\usepackage[math]{cellspace}

\usepackage{lipsum}
\usepackage{comment}
\usepackage{soul}
\usepackage[normalem]{ulem}

\usepackage[colorlinks,
linkcolor=BrickRed,
citecolor=MidnightBlue,
urlcolor=MidnightBlue,
bookmarks=true,        
bookmarksopen=true,    
bookmarksnumbered=true,
]{hyperref}

\renewcommand{\i}{\mathrm{i}}

\newcounter{ls}

\newcounter{pr}

\newcounter{kw}

\newcounter{qg}

\begin{document}
	
\title{Experimental Detection of Dissipative Quantum Chaos}

\author{Kristian Wold}\thanks{These authors contributed equally}
\affiliation{Department of Computer Science, OsloMet – Oslo Metropolitan University, N-0130 Oslo, Norway\looseness=-1}

\author{Zitian Zhu}\thanks{These authors contributed equally}
\affiliation{School of Physics, ZJU-Hangzhou Global Scientific and Technological Innovation Center, and Zhejiang Key Laboratory of Micro-nano Quantum Chips and Quantum Control, Zhejiang University, Hangzhou, China}

\author{Feitong Jin}
\author{Xuhao Zhu}
\author{Zehang Bao}
\author{Jiarun Zhong}
\author{Fanhao Shen}
\author{Pengfei Zhang}
\author{Hekang Li}
\author{Zhen Wang}
\author{Chao Song}
\author{Qiujiang Guo}
\email{qguo@zju.edu.cn}
\affiliation{School of Physics, ZJU-Hangzhou Global Scientific and Technological Innovation Center, and Zhejiang Key Laboratory of Micro-nano Quantum Chips and Quantum Control, Zhejiang University, Hangzhou, China}

\author{Sergey Denisov}
\email{sergiyde@oslomet.no}
\affiliation{Department of Computer Science, OsloMet – Oslo Metropolitan University, N-0130 Oslo, Norway\looseness=-1}

\author{Lucas  S\'a}
\email{ld710@cam.ac.uk}
\affiliation{TCM Group, Cavendish Laboratory, Ray Dolby Centre, University of Cambridge, JJ Thomson Avenue, Cambridge, CB3 0US UK}

\author{H. Wang}
\affiliation{School of Physics, ZJU-Hangzhou Global Scientific and Technological Innovation Center, and Zhejiang Key Laboratory of Micro-nano Quantum Chips and Quantum Control, Zhejiang University, Hangzhou, China}

\author{Pedro Ribeiro}
\email{ribeiro.pedro@tecnico.ulisboa.pt}
\affiliation{CeFEMA, Instituto Superior T\'ecnico, Universidade de Lisboa, Av.\ Rovisco Pais, 1049-001 Lisboa, Portugal}
\affiliation{Beijing Computational Science Research Center, Beijing 100193, China}

\begin{abstract}
More than four decades of research on chaos in isolated quantum systems have led to the identification of universal signatures -- such as level repulsion and eigenstate thermalization -- that serve as cornerstones in our understanding of complex quantum dynamics.
The emerging field of dissipative quantum chaos explores how these properties manifest in open quantum systems, where interactions with the environment play an essential role.
We report the first experimental detection of dissipative quantum chaos and integrability by measuring the complex spacing ratios (CSRs) of open many-body quantum systems implemented on a high-fidelity superconducting quantum processor.
Employing gradient-based tomography, we retrieve a ``donut-shaped'' CSR distribution for chaotic dissipative circuits, a hallmark of level repulsion in open quantum systems. For an integrable circuit, spectral correlations vanish, evidenced by a sharp peak at the origin in the CSR distribution. 
As we increase the depth of the integrable dissipative circuit, the CSR distribution undergoes an integrability-to-chaos crossover, demonstrating that intrinsic noise in the quantum processor is a dissipative chaotic process.
Our results reveal the universal spectral features of dissipative many-body systems and establish present-day quantum computation platforms, which are predominantly used to run unitary simulations, as testbeds to explore dissipative many-body phenomena.
\end{abstract}

\maketitle

Quantum computers are a disruptive tool to tackle many-body physics and quantum chemistry problems~\cite{Feynman1982,altman2021PRXQ}, while also enabling more general-purpose algorithms that harness quantum superposition and entanglement -- most notably, the quantum Fourier transform for efficient integer factoring~\cite{Shor1994}. 
These capabilities set quantum computers to address computational challenges beyond the reach of classical machines~\cite{NielsenChuang2010}.
However, one of the central hurdles to achieving robust quantum computation is noise~\cite{preskill2018}.
Noise inevitably arises from interactions with the environment, leading to dissipation and decoherence effects~\cite{breuerbook}. 
While they have often been viewed as imperfections to be mitigated, these effects also create an arena for exploring new regimes of quantum chaos (QC).

Early work on QC in the 1970s -- 80s focused on unitary systems, with pioneering analysis of nuclear spectra~\cite{haq-pandey-bohigas_1982,bohigas-pandey-haq_1983} and experiments with microwave billiards~\cite{Bohigas1984,Stockmann1990,Stockmann1999}. The latter offered controllable platforms for testing theoretical predictions and contributed to uncovering deep connections between classical chaos, spectral statistics of quantum Hamiltonians, and their universal symmetries~\cite{Brody1981,guhr1998,Haake2010}. 
However, extending the QC framework to more general settings -- such as dissipative quantum systems, whose degrees of freedom interact with the surrounding environment -- introduces additional layers of complexity.
Over the past five years, there have been significant theoretical advances in the study of the dynamics of dissipative quantum systems, and random matrix theory (RMT)~\cite{Mehta2004,Haake2010} has proven to be an invaluable tool for characterizing the complex spectra of such systems~\cite{Denisov2019, Can2019, Sa2020JPA}, helping to uncover universal features that herald chaotic behavior~\cite{Sa2020PRX,sa2022PRX,sa2023PRX,kawabata2023PRXQ}. Despite this progress, the phenomenon of dissipative quantum chaos (DQC), in which unitary and dissipative dynamics intricately intertwine, has so far remained a largely theoretical concept, despite growing interest in its possible manifestations in quantum technologies. 

Here, we report the first experimental observation of DQC using a state-of-the-art quantum processor.
Similarly to the role that the experiments with microwave billiards played in observing and controlling Hamiltonian QC~\cite{Stockmann1990,Stockmann1999}, 
high-quality superconducting platforms enable studies and control of dissipative quantum many-body processes. 
By tuning dissipation in the quantum circuits, we detect signatures of both chaotic and integrable regimes. We also show that the hardware's intrinsic dissipative processes are fully chaotic, allowing DQC to eventually dominate the dynamics in sufficiently deep circuits.
By demonstrating that it is experimentally feasible to uncover the spectral features diagnosing dissipative chaotic behavior, we establish present-day quantum processors as flexible platforms for testing theoretical models and expanding our understanding of complex dissipative quantum systems.
Our findings deepen the understanding of dissipative quantum phenomena and offer new ideas for harnessing quantum chaos.

\begin{figure*}[t]
     \centering
     \includegraphics[width=\textwidth]{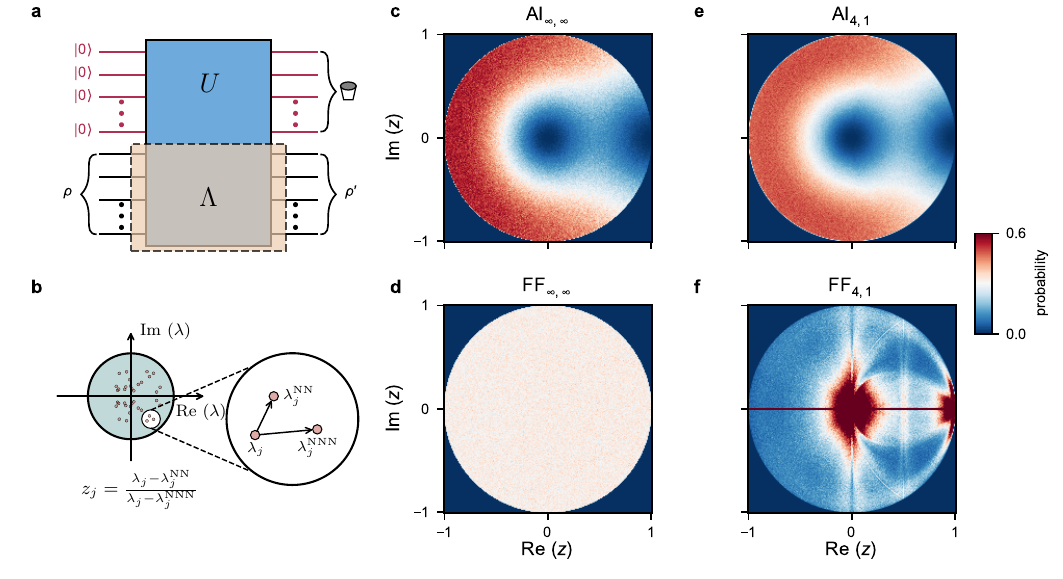}
     \caption{
     \label{fig:donut}
     \textbf{Dissipative quantum chaos and integrability in infinite and finite systems.}
     \textbf{\textsf{a}}, Quantum map $\Lambda$ obtained by evolving the system and ancilla qubits with a unitary circuit $U$, after which the ancilla qubits are discarded. The system qubits are prepared in an initial state $\rho$ and evolve into a final state $\rho^{\prime}$, while the ancilla qubits are initialized in a fixed pure state. 
     \textbf{\textsf{b}}, Computation of the complex spacing ratios  (CSRs) for the spectrum of $\Lambda$. We illustrate the computation of $z_j$ from the eigenvalue $\lambda_j$ and its nearest ($\lambda_j^{\text{NN}}$) and next-to-nearest ($\lambda_j^{\text{NNN}}$) neighbors, which is then repeated for the whole spectrum.
     \textbf{\textsf{c}--\textsf{f}}, Numerical simulations of the CSR distribution in the complex plane for chaotic and integrable systems of different sizes. For clarity, all histogram plots have been truncated at a maximum probability value of 0.6.
     \textbf{\textsf{c}}, AI$_{\infty,\infty}$: Approximation to the distribution for infinitely-large random matrices in the symmetry class AI, computed by exact diagonalization of $256$ $2^{15}\times2^{15}$ random GinUE matrices~\cite{ginibre1965}, for which the CSR distribution has approximately converged to the thermodynamic limit~\cite{Sa2020PRX,sa2022PRX}. 
     \textbf{\textsf{d}},  FF$_{\infty,\infty}$: Distribution for 256 samples of $2^{15}$ uncorrelated complex normal random variables, as an approximation to the distribution for infinite systems.
     \textbf{\textsf{e}}, AI$_{4,1}$: Distribution for an ensemble of $10^5$ quantum maps obtained by tracing out the ancilla qubit of a Haar-random $U$ of the same dimension as in the experimental chaotic circuit ($4$ physical qubits and $1$ ancilla qubit), cf.\ Fig.~\ref{fig:circuit}d. The characteristic ``donut shape'' is preserved at finite sizes.
     \textbf{\textsf{f}}, FF$_{4,1}$: Distribution for the steady-state sector of an ensemble of $10^5$ quantum maps obtained by tracing out the ancilla qubit of a FF $U$ with the same size and symmetries as the experimental integrable circuit (cf.\ Fig.~\ref{fig:circuit}c), constructed as described in Supplementary Information~\ref{app:numerical_implementation}. There are strong finite-size corrections to the distribution, namely, a peak near the origin superimposed on the flat distribution.}
\end{figure*}

\begin{figure*}[t]
     \centering
     \includegraphics[width=0.75\textwidth]{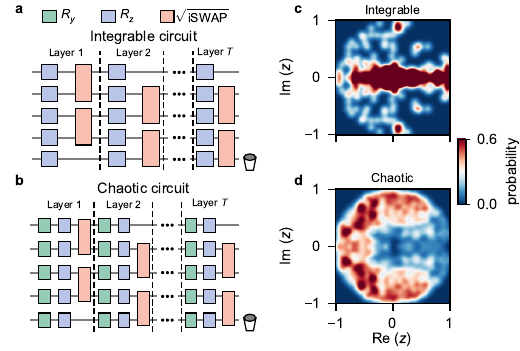}
     \caption{
     \textbf{Measured complex spacing ratio (CSR) distributions for finite-size dissipative quantum chaos and integrability.}
     A 4-qubit quantum map is generated by applying a 5-qubit circuit with $T$ layers and tracing out an ancilla qubit. In the experiments, the circuits are further compiled into two-qubit CZ gates and single-qubit gates that can be naturally realized on the processor (see Supplementary Information~\ref{SM_ExpDetail}). 
     \textbf{\textsf{a}}, Integrable (free-fermion) circuit. Blue gates implement a $Z$-rotation with an independently sampled random angle for each site and layer, while $\sqrt{\mathrm{iSWAP}}$ gates connect neighboring qubits in an alternating brickwork fashion. 
     \textbf{\textsf{b}}, The chaotic dissipative circuit follows the same structure as the integrable one, with additional $Y$ rotations with randomly sampled angles at the beginning of each layer.
     \textbf{\textsf{c}}, Measured CSR distribution for an ensemble of 7 integrable circuits with $T=5$. For this depth, particle number is approximately conserved (see Supplementary Information~\ref{app:approximate_symmetries}) and we consider the steady-state sector only.
     \textbf{\textsf{d}}, Measured CSR distribution for an ensemble of 10 chaotic circuits with $T=10$. For this depth, the unitary circuit (i.e., before tracing out the ancillary qubit) is effectively indistinguishable from a Haar-random unitary~\cite{Wold2024}. Gaussian smearing with standard deviation $\delta = 0.05$ is applied to the experimental results in \textbf{\textsf{c}} and \textbf{\textsf{d}}  to allow the comparison with the theoretical predictions in Fig.~\ref{fig:donut}. The histograms have been truncated at a maximum probability value of 0.6.
     }
     \label{fig:circuit}
\end{figure*}

\vspace{.4cm}
\noindent\textbf{\large Integrable and chaotic quantum maps}

\noindent A convenient way to describe the evolution of a quantum state in an open quantum system is through a quantum map, $\Lambda$, which transforms the density matrix of the state, $\rho$, into a new density matrix, $\rho' = \Lambda(\rho)$~\cite{NielsenChuang2010}; see  Fig.~\ref{fig:donut}a. 
A practical realization of the map $\Lambda$ involves applying a unitary evolution that couples the system to ancillary environment qubits, followed by discarding (or ‘forgetting’) the information contained in the environment qubits; see Fig.~\ref{fig:donut}a and Supplementary Information~\ref{app:numerical_implementation}.
Such a procedure mimics the effect of an environment, which continuously interacts with the system and effectively measures and discards certain degrees of freedom.

The regular or chaotic nature of dissipative quantum processes can be characterized by the spectral properties of  $\Lambda$~\cite{grobe1988,akemann2019,Sa2020PRX}. In chaotic processes, the map’s complex eigenvalues, $\lambda_{i}$ ($i=1,2,\dots$), exhibit repulsion, reflecting the behavior of random matrices. In contrast,  eigenvalues of regular or integrable maps do not display this feature. Complex spacing ratios (CSR)~\cite{Sa2020PRX} offer a direct way to quantify the repulsion effect by examining statistics of 
\[
z_i = \frac{\lambda_i - \lambda_i^{\text{NN}}}{\lambda_i - \lambda_i^{\text{NNN}}},
\]
where $\lambda_i^{\text{NN}}$ and $\lambda_i^{\text{NNN}}$ are the nearest- and next-to-nearest neighbors of the eigenvalue $\lambda_i$; see Fig.~\ref{fig:donut}b. Note that purely real eigenvalues should be excluded from the analysis, since they have statistics different from the statistics of the bulk eigenvalues.
For an ensemble of infinite-size chaotic maps, the spectral correlations are in agreement with those of a non-Hermitian real random matrix (known as the Ginibre orthogonal ensemble~\cite{ginibre1965} or the AI class~\cite{kawabata2019PRX,hamazaki2022,Sa2020PRX}). 
The distribution $P(z)$ in the complex plane for infinite (AI$_{\infty,\infty}$) and finite (AI$_{4,1}$) systems is shown in Fig.~\ref{fig:donut}c, e, respectively, and takes on a characteristic ``bitten donut'' shape: level repulsion suppresses the density near both the center and the positive real axis.
On the other hand, uncorrelated eigenvalues of integrable free-fermion (FF) systems yield a flat ``pie-like'' distribution over the unit disk (corresponding to planar Poisson statistics~\cite{Sa2020PRX}), see Fig.~\ref{fig:donut}d, indicating the absence of level repulsion.

To observe dissipative quantum chaos and its breakdown using a superconducting qubit processor, we construct 
digital quantum circuits that can effectively emulate integrable or chaotic dynamics. As shown in Fig.~\ref{fig:circuit}a, we implement a 5-qubit unitary non-chaotic quantum circuit and realize a 4-qubit dissipative quantum map by tracing out an ancilla qubit.
The unitary dynamics consists of $T$ layers of $Z$ rotations with randomly sampled angles and $\sqrt{\mathrm{iSWAP}}$ gates connecting neighboring qubits in an alternating brickwork fashion. These gates form a subset of so-called matchgates, known as free-fermion gates in the many-body physics community~\cite{Jozsa2008}. In this family of models, integrability remains intact even after tracing over the ancilla qubit. 
We break the integrability of the free-fermion circuit by inserting an additional set of $Y$ rotations in each layer; see Fig.~\ref{fig:circuit}b.
For $T \geq 10$, the unitary evolution generated by this circuit is effectively indistinguishable from a Haar-random 5-qubit unitary~\cite{Wold2024} (i.e., a $2^5 \times 2^5 $ matrix sampled uniformly from the unitary group).

\begin{figure*}[t]
     \centering
     \includegraphics[width=\textwidth]{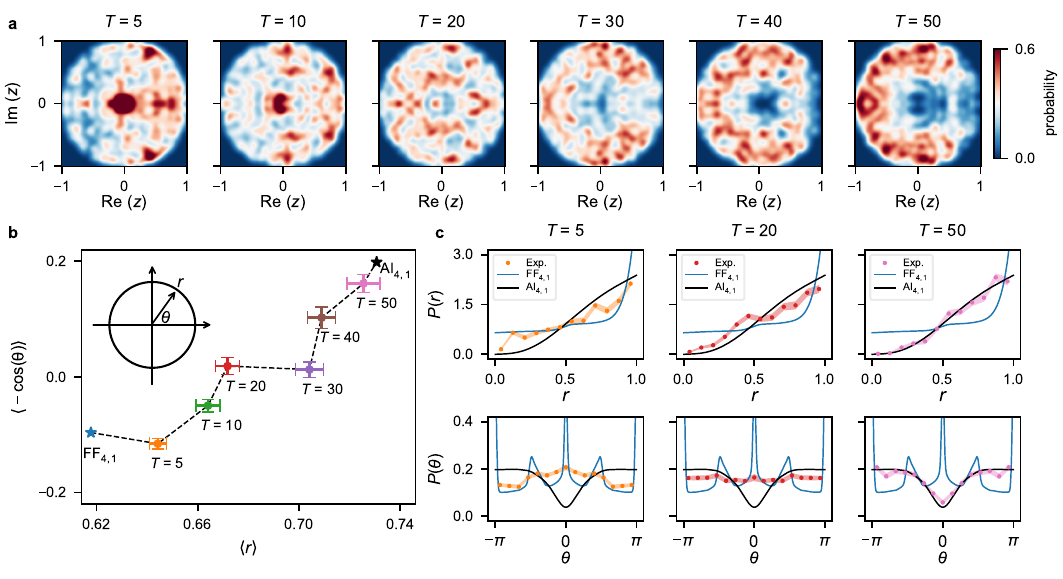}
     \caption{\textbf{Integrability-to-chaos transition with increasing circuit depth.} 
     \textbf{\textsf{a}}, Complex spacing ratio distribution $P(z)$ for an ensemble of $10$ integrable circuits of different depth $T$. The distribution exhibits a characteristic free-fermion peak at the origin for $T = 5$, evolves into an approximately flat disk by $T = 20$, and eventually develops the bitten-donut shape that is indicative of quantum chaos at $T = 50$. Particle number is strongly broken at large $T$ (see Supplemental information~\ref{app:approximate_symmetries}), hence contrary to Fig.~\ref{fig:circuit} we do not separate different symmetry sectors.
     \textbf{\textsf{b}}, Scatter plot of $\langle r \rangle$ vs $\langle -\cos(\theta) \rangle$, where $z=r e^{i\theta}$. We compare the experimental results with simulations of ideal AI$_{4,1}$ and FF$_{4,1}$ statistics (also without separating symmetry sectors). As $T$ increases the experimental points crossover from FF$_{4,1}$ (integrability) to AI$_{4,1}$ (chaos). Error bars are estimated by bootstrap resampling of the 10 experimental realizations (see Supplemental information~\ref{app:BootstrapErrorEstimation}).
     \textbf{\textsf{c}}, Marginal radial (top), $P(r)$, and angular (bottom), $P(\theta)$, distributions of $z$. We compare the experimental results with the simulations of the ideal statistics for AI$_{4,1}$ (chaotic) and FF$_{4,1}$ (integrable).} 
    \label{fig:integrable_vs_chaos}
\end{figure*}

To obtain accurate finite-size predictions for comparison with experimental data, we computed the CSR distributions for generic chaotic and free-fermion quantum maps with $4$ system and $1$ environment qubits, denoted AI$_{4,1}$ and FF$_{4,1}$, respectively (see Supplementary Information~\ref{app:numerical_implementation}).
Figure~\ref{fig:donut}e--f show the simulated CSR distribution $P(z)$. Due to finite-size effects, these distributions deviate from the theoretical predictions in Fig.~\ref{fig:donut}c--d, which are based on large random matrix models that approximate the thermodynamic limit, where the number of qubits tends to infinity.
The most notable deviations are the line-like features in $P(z)$ and a pronounced peak at the origin observed for the free-fermion maps. 
A scaling analysis confirms this is a finite-size distortion and recovers the characteristic pie-like CSR structure of integrable systems in the thermodynamic limit (see Supplementary Information~\ref{app:finite-size}); nevertheless, this distortion makes the CSR distribution even more distinct from the chaotic case. 
Moreover, the free-fermion circuits under consideration conserve ``particle number'', i.e., the difference between the number of logical $1$s and $0$s, resulting in a spectrum of $\Lambda$ that splits into independent sectors labeled by fixed quantum numbers. For a meaningful comparison with uncorrelated eigenvalues and random matrix theory, the statistical analysis must be performed in each sector separately~\cite{guhr1998}.
This is so because a CSR analysis of all eigenvalues cannot independently prove integrability: the superposition of independent random-matrix spectra, characteristic of chaotic systems with symmetries, also displays Poisson statistics.
We compute the CSR for the steady-state sector (``half-filling'' sector); see Supplementary Information~\ref{app:FF_Symmetries} for details.

To experimentally measure CSR distributions and verify theoretical predictions, we implemented the circuits of Fig.~\ref{fig:circuit} on our quantum processor, which is based on high-quality superconducting qubits with average energy relaxation time of $\sim$150 $\mu$s, single-qubit gate fidelity of $\sim99.98\%$, and two-qubit CZ gate fidelity of $\sim99.86\%$ (see Supplementary Information~\ref{SM_ExpDetail}). We performed a gradient-based process tomography on four qubits, including a full treatment of state preparation and measurement (SPAM) errors, to retrieve the resulting quantum map; details of the algorithm can be found in Ref.~\cite{Wold2024} and Supplementary Information~\ref{app:MapRetrival}. We then diagonalized the retrieved map to obtain its spectrum and computed the CSR distribution. The results are shown in Fig.~\ref{fig:circuit}c--d.
Experimentally, particle number is approximately conserved for $T=5$ (see Supplementary Information~\ref{app:approximate_symmetries}), hence we considered the steady-state sector only.
Our results show remarkable agreement with the finite-size simulated data. In particular, level repulsion is clearly absent in the integrable case (Fig.~\ref{fig:circuit}d), while the depletion near the origin in Fig.~\ref{fig:circuit}c matches the expected behavior for correlated eigenvalues in the chaotic regime.

\vspace{.4cm}
\noindent\textbf{\large Intrinsic noise as a dissipative quantum chaotic process}

\noindent Having established protocols for engineering and retrieving chaotic and integrable dissipative circuits, we now address the critical role of intrinsic hardware noise. While error rates appear modest in individual operations (0.029\% per single-qubit gate and 0.176\% per CZ gate), we demonstrate how cumulative noise effects drive a crossover from integrable to chaotic dissipative dynamics.

We probe this transition by measuring the CSR distribution of integrable circuits of increasing depth $T \in [5,50]$.
Since generic noise inherently breaks the engineered particle-number symmetry, we do not resolve symmetry sectors. 
Figure~\ref{fig:integrable_vs_chaos}a reveals three distinct regimes as follows. 
\textit{Shallow circuits} ($T \leq 10$): Preservation of the CSR peak near the origin, consistent with residual integrability; 
\textit{Crossover regime} ($10 < T \leq 40$): Gradual spectral redistribution from a peaked to a ``donut-shaped'' distribution, signaling the onset of dissipative quantum chaos;
\textit{Deep chaotic phase} ($T > 40$): Fully-developed dissipative quantum chaos, with the same spectral statistics as a random matrix in class AI.

To quantitatively measure the degree of chaos of the quantum map, we compare the marginal distributions $P(r)=\int P(z) r\, \mathrm{d}\theta$ and $P(\theta)=\int P(z) r\, \mathrm{d}r$ of the CSR variable $z=r e^{i \theta}$ for different $T$. Figure~\ref{fig:integrable_vs_chaos}c shows the transition from integrable to chaotic behavior as the depth increases. 
The $\left\langle r \right\rangle$ vs $\left\langle \cos\theta \right\rangle$ plane, Fig.~\ref{fig:integrable_vs_chaos}b, summarizes these findings.
The convergence of experimental data to AI$_{4,1}$ with increasing circuit depth indicates the growing dominance of the chaotic component of the quantum map, demonstrating hardware noise as a source of dissipative quantum chaos.

\vspace{.4cm}
\noindent\textbf{\large Discussion}

\noindent We presented the first experimental realization of dissipative quantum chaos, a milestone for open quantum systems. By harnessing a modern quantum processor, we engineered and analyzed integrable and chaotic dissipative quantum maps. Noise, often seen as a hindrance, emerges here as a tool to probe chaos in sufficiently deep circuits.

Our approach experimentally reveals that the characteristic level repulsion observed through the CSR distributions associated with chaos contrasts sharply with the uncorrelated spectra of integrable maps. As circuit depth increases, accumulated noise disrupts the engineered symmetries, leading to fully chaotic behavior. This transition demonstrates the pivotal role of dissipation in shaping quantum dynamics.

By showing that dissipative quantum chaos is experimentally accessible, we open new avenues for advancing theoretical frameworks and real-world quantum technologies. Although shallow circuits can retain integrability, deeper circuits inevitably encounter chaotic effects. Understanding this transition is crucial for error mitigation and quantum error correction. It also underscores the importance of robust strategies for quantum simulation, as chaos ultimately affects computational outcomes.

Our results suggest a new operational metric for benchmarking quantum processing platforms based on their susceptibility to dissipative chaos. By implementing an integrable dissipative circuit and monitoring the CSR statistics as a function of circuit depth, one can extract a characteristic crossover time at which the spectral correlations transition from integrable to chaotic. This timescale is notably shorter than both the coherence and relaxation times, providing a sensitive indicator of intrinsic noise that complements existing metrics.

The robustness and flexibility of our method enable further exciting possibilities, including studies of dissipative many-body localization, phase transitions in open quantum systems, and different types of symmetries in quantum channels. Our work thus opens the door to the new field of experimental dissipative quantum chaos.\\

\vspace{.4cm}
\noindent\textbf{Acknowledgements}\\
The experiment was implemented on the superconducting quantum platform in Zhejiang University.
The device was fabricated at the Micro-Nano Fabrication Center of Zhejiang University.
The experimental team acknowledges the support from the National Natural Science Foundation of China (Grant Nos. 12274368, 12174342, 12274367, 12322414, 12404570, 12404574, 92365301), the Zhejiang Provincial Natural Science Foundation of China (Grant Nos. LR24A040002 and LDQ23A040001), and the National Key Research and Development Program of China (Grant No. 2023YFB4502600). 
This work was supported by the Research Council of Norway, project ``IKTPLUSS-IKT og digital innovasjon
- 333979'' (KW and SD), and by FCT-Portugal, Grant Agreement No. 101017733 (PR), as part of the QuantERA II project “DQUANT: A Dissipative Quantum Chaos perspective on Near-Term Quantum Computing”~\footnote{\url{https://doi.org/10.54499/QuantERA/0003/2021}}. 
PR acknowledges further support from FCT through the financing of the I\&D unit: UID/04540 - Centro de Física e Engenharia de Materiais Avançados. 
LS was supported by a Research Fellowship from the Royal Commission for the Exhibition of 1851.

\vspace{.4cm}
\noindent\textbf{Author Contributions}\\
LS, SD, PR, and KW conceived the project. 
ZZ, FJ, and XZ conducted the experiments under the supervision of QG and HW.
HL fabricated the device under the supervision of HW.
HW, QG, CS, ZW, HL, PZ, FS, JZ, ZB, XZ, FJ, and ZZ contributed to the experimental setup.
PR, LS, SD, and KW developed the theoretical framework, performed simulations, and analyzed the experimental data. 
LS, PR, KW, SD, QG, and ZZ wrote the manuscript with input from all authors. All authors discussed the results and contributed to the final version of the manuscript.

\vspace{.4cm}
\noindent\textbf{Competing interests}\\
The authors declare no competing interests.

\vspace{.4cm}
\noindent\textbf{Supplementary Information}\\
Supplementary Information is available for this paper.

\vspace{.4cm}
\noindent\textbf{Data availability}\\
The data presented in the figures and that support the other findings of this study will be publicly available upon its publication. 

\vspace{.4cm}
\noindent\textbf{Code availability}\\
All the relevant source codes are available from the corresponding authors upon reasonable request. 

\bibliography{main_v1}


\clearpage

\setcounter{table}{0}
\renewcommand{\thetable}{S\arabic{table}}%
\setcounter{figure}{0}
\renewcommand{\thefigure}{S\arabic{figure}}%
\setcounter{equation}{0}
\renewcommand{\theequation}{S\arabic{equation}}%
\setcounter{page}{1}
\renewcommand{\thepage}{SI-\arabic{page}}%
\setcounter{secnumdepth}{3}
\setcounter{section}{0}
\renewcommand{\thesection}{S\Roman{section}}%
\setcounter{subsection}{0}
\renewcommand{\thesubsection}{\arabic{subsection}}%

\onecolumngrid
\begin{center}\large{
		\textbf{Supplementary Information for}\\
		\textbf{Experimental Detection of Dissipative Quantum Chaos}\vspace{2ex}
        }
\end{center}
\twocolumngrid

\section{Experimental details}
\label{SM_ExpDetail}
\subsection{Device information}
\begin{figure}[h]
    \begin{center}
    \includegraphics[width=1\columnwidth]{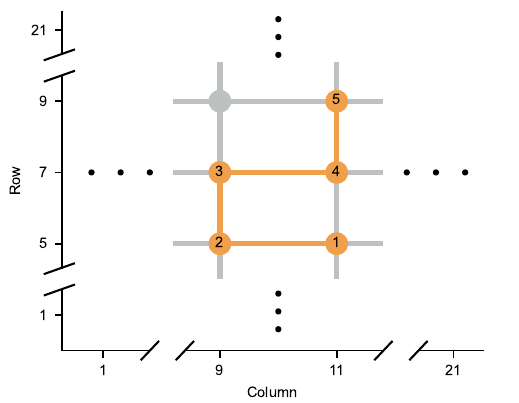}
    \end{center}
    \caption{{\bf Schematic of the five-qubit chain arrangement on the quantum processor.}
    There are 121 qubits (circles) and 220 tunable couplers (lines) on the processor, from which we choose 5 qubits and 4 couplers to perform the experiment. They are highlighted in orange in the diagram.}
    \label{smfig:layout}
\end{figure}

\begin{figure}[h!]
    \begin{center}
    \includegraphics[width=1\columnwidth]{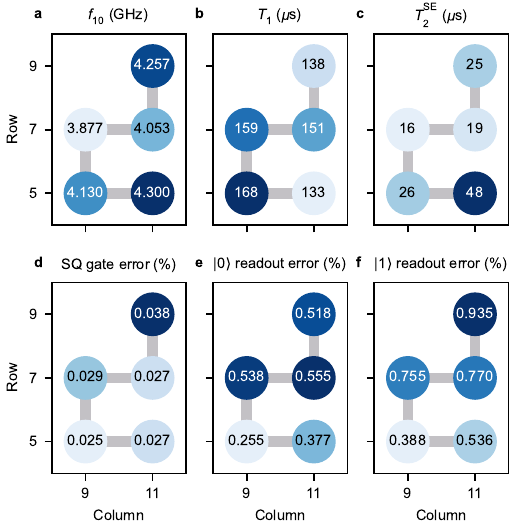}
    \end{center}
    \caption{{\bf Heatmaps of typical parameters of the five qubits used in our experiments.}
    {\bf \textsf{a}}, Qubit idle frequency $f_{10}$.
    {\bf \textsf{b}}, Qubit energy relaxation time $T_1$.
    {\bf \textsf{c}}, Qubit spin-echo dephasing time $T_2^{\rm SE}$.
    {\bf \textsf{d}}, Single-qubit (SQ) gate Pauli error.
    {\bf \textsf{e}}, Qubit $|0\rangle$ state readout error.
    {\bf \textsf{f}}, Qubit $|1\rangle$ state readout error.
    }
    \label{smfig:heatmap}
\end{figure}

\begin{figure}[h!]
    \begin{center}
    \includegraphics[width=1\columnwidth]{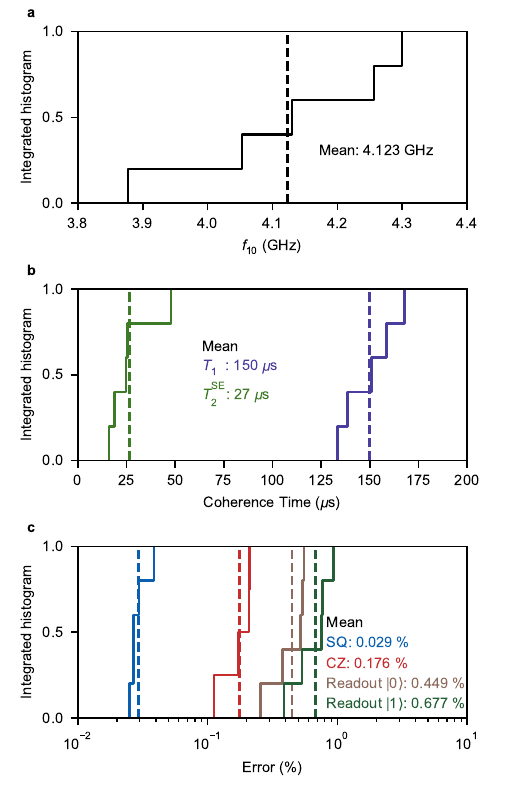}
    \end{center}
    \caption{{\bf Integrated histograms of qubit performance.}
    {\bf \textsf{a}}, Distribution of qubit idle frequency $f_{10}$.
    {\bf \textsf{b}}, Distribution of qubit energy relaxation time $T_1$ (purple line) and spin-echo dephasing time $T_2^{\rm{SE}}$ (green line).
    {\bf \textsf{c}}, Distribution of gate and readout errors. The blue, red, brown, and green lines represent single-qubit gate errors (SQ), two-qubit CZ gate errors (CZ), and $|0\rangle$ and $|1\rangle$ readout errors, respectively. The dashed lines indicate their average values. The gate errors are measured with simultaneous cross-entropy benchmarking (XEB)~\cite{Boixo2018NP}.}
    \label{smfig:device}
\end{figure}

The five-qubit chain (Fig.~\ref{smfig:layout}) utilized in the experiments is selected on an $11\times 11$ superconducting quantum processor~\cite{xu2023digital,Bao2024NC}. Figure~\ref{smfig:heatmap} displays the basic performance parameters of the five qubits and Fig.~\ref{smfig:device} shows the corresponding integrated histograms. The mean qubit idle frequency is about 4.123 GHz (Fig.~\ref{smfig:device}a). The mean value of qubit energy relaxation time $T_1$ and spin-echo dephasing time $T_2^{\rm SE}$ are $\sim$150 $\mu$s and $\sim$27 $\mu$s, respectively (Fig.~\ref{smfig:device}b). Figure~\ref{smfig:device}c shows the distribution of single-qubit (SQ) and two-qubit CZ gate Pauli errors, with mean values of $0.029 \%$ and $0.176\%$, respectively. The mean $|0\rangle$ and $|1\rangle$ readout errors are 0.449\% and 0.677\%, respectively.

\subsection{Experimental circuit optimization}\label{app:Experimental circuit optimization}
\begin{figure}[h!]
    \begin{center}
    \includegraphics[width=1\columnwidth]{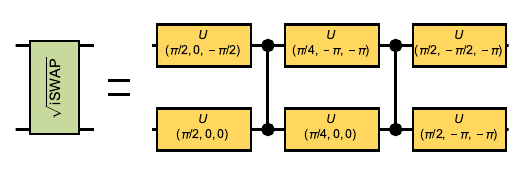}
    \end{center}
    \caption{{\bf $\boldsymbol{\sqrt{\rm iSWAP}}$ gate decomposition.}
    The $\sqrt{\rm iSWAP}$ gate is decomposed into six Qiskit $U$ gates and two CZ gates, which can be realized in our experiment conveniently.
    }
    \label{smfig:sqrt_iswap}
\end{figure}

Figure~\ref{fig:circuit}a--b of the main text shows the theoretical circuits of dissipative quantum chaos and integrability. In our experiment, we further transpile them into circuits that only contain single-qubit and CZ gates using Qiskit~\cite{qiskit2024}. For example, as illustrated in Fig.~\ref{smfig:sqrt_iswap}, the $\sqrt{\rm iSWAP}$ gate can be decomposed into the combinations of Qiskit $U$ gates and CZ gates, where
\begin{align}
    \sqrt{\rm iSWAP}&= \begin{pmatrix}
        1&0&0&0\\
        0&\frac{1}{\sqrt{2}}&\frac{1}{\sqrt{2}}i&0\\
        0&\frac{1}{\sqrt{2}}i&\frac{1}{\sqrt{2}}&0\\
        0&0&0&1
    \end{pmatrix}, \\
    U(\theta, \phi, \lambda) &=
    \begin{pmatrix}
        \cos\left(\theta/2\right) & -e^{i\lambda}\sin\left(\theta/2\right) \\
        e^{i\phi}\sin\left(\theta/2\right) & e^{i(\phi+\lambda)}\cos\left(\theta/2\right)
    \end{pmatrix},\\
    {\rm CZ}&= \begin{pmatrix}
        1&0&0&0\\
        0&1&0&0\\
        0&0&1&0\\
        0&0&0&-1
    \end{pmatrix}.
\end{align}

\noindent Such circuits can be executed on our processor directly because we can naturally realize arbitrary single-qubit rotation and high-fidelity CZ gates on our experimental platform.

\begin{figure}[h]
    \begin{center}
    \includegraphics[width=1\columnwidth]{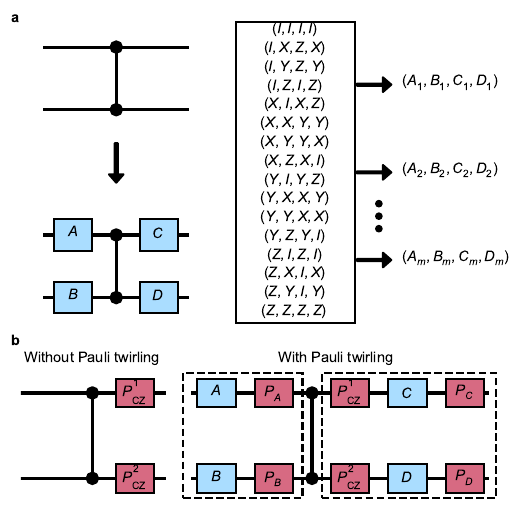}
    \end{center}
    \caption{{\bf Pauli twirling technique and possible combinations.}
    {\bf \textsf{a}}, We wrap each CZ gate in the circuit with four single-qubit gates $A, B, C$ and $D$, which are randomly sampled from the 16 possible combinations listed in the box. Ideally, the twirled CZ remains unchanged. However, in the experiment, coherent noise is unavoidable. It can be helpful to convert such noise to stochastic noise by executing the experimental circuit with different random Pauli twirling realizations $m$ times and averaging the measurement outcomes.
    {\bf \textsf{b}}, Noisy CZ gates without (left) and with (right) Pauli twirling. $P^i_j$ denotes noisy gates. For the CZ gate without Pauli twirling, $P_{\rm CZ}^i$ can be coherent and the errors accumulate each time we apply the CZ gate. The Pauli twirling introduces randomness to the circuit to make noisy gates cancel each other out or become incoherent.
    }
    \label{smfig:twirling}
\end{figure}

We employ Pauli twirling~\cite{Wallman2016Twirling} to suppress nonstochastic 
noise~\cite{Li2017Twirling} in the experiment. Specifically, as illustrated in Fig.~\ref{smfig:twirling}a, we randomly choose four single-qubit gates from the 16 possible combinations and wrap CZ with these gates. $m$ denotes the number of random circuit realizations, which is 6 in our experiment. For each realization, we do 2000 single-shot measurements. The coherent noise is suppressed by averaging over the measurement outcomes of $m$ realizations. We can understand it intuitively through the depolarizing error model~\cite{arute2019quantum,xiang2024long}. That is, for each single-qubit and CZ gate, errors can occur with an error rate $e_p$. We can simulate such an error for a $d$-qubit gate by applying a randomly chosen Pauli string $P\in\{I,X, Y, Z\}^{\otimes d}$ ($P\neq I^{\otimes d}$) with probability $e_p/(4^d-1)$. As depicted in Fig.~\ref{smfig:twirling}b, the left panel is a regular CZ gate followed by noisy gates $P_{\rm CZ}^1 \otimes P_{\rm CZ}^2$ and the right panel displays twirled CZ with corresponding noisy gates. Due to the randomness of the gates in the two dashed boxes, they may cancel each other out to suppress noise. More typically, the noise can be coherent, which means $P_{\rm CZ}^1$ and $P_{\rm CZ}^2$ are not random but biased; errors will accumulate in such a scenario. Pauli twirling destroys such coherent noise, thus protecting the quantum circuits.

\section{Numerical details}
\subsection{Map retrieval procedure}
\label{app:MapRetrival}

To retrieve quantum maps that model noisy quantum circuits, we utilize the gradient-based quantum process tomography protocol introduced in Ref.~\cite{Wold2024}. The protocol implements a parameterized Kraus map
\begin{equation}\label{eq:map}  
 \mathcal{T}_r(\rho;\boldsymbol{\theta}) = \sum_{s=1}^{r}K_s(\boldsymbol{\theta})\rho K^{\dagger}_s(\boldsymbol{\theta}),~ 
    \sum_{s=1}^{r} K^{\dagger}_s(\boldsymbol{\theta})K_s(\boldsymbol{\theta}) = I_{d},
\end{equation}
where $\boldsymbol{\theta}$ is a parameter vector,  
$K_s(\boldsymbol{\theta}) \in \mathbb{C}^{d\times d}$ are Kraus operators and $1 \leqslant r  \leqslant d^2$ is the rank of the map. To model general quantum maps, we utilize full rank, $r = d^2$, in this work. The second identity in Eq.~\eqref{eq:map} imposes the trace-preservation condition, which is ensured for any choice of parameters $\boldsymbol{\theta}$.

Since our retrieval procedure is based on tomography data, we need to account for state preparation and measurement (SPAM) errors. 
We do this by considering a non-ideal initial density operator 
\begin{equation}\label{eq:init}
    \rho_0 = \frac{A_{\rho}A_{\rho}^{\dagger}}{\text{Tr}[A_{\rho}A_{\rho}^{\dagger}]},
\end{equation} parameterized by a complex $d \times d$ matrix $A_{\rho}$. To account for measurement errors, we can model positive operator-valued measure (POVM) elements in the most general case by using Eq.~(\ref{eq:map}), and identify
\begin{equation}\label{eq:POVM}
    E_j(\boldsymbol{\omega}) = K_j^\dagger(\boldsymbol{\omega}) K_j(\boldsymbol{\omega}).
\end{equation} However, we instead focus on the more special case of readout errors, utilizing a $d \times d$ stochastic corruption matrix
$C$. This matrix is parameterized by a $d \times d$ real matrix $A_\text{C}$, 
\begin{equation}\label{eq:CM}
    C_{jl} = \frac{|(A_\text{C})_{jl}|}{\sum_{k=1}^d |(A_\text{C})_{kl}|}.
\end{equation} 
$C_{ij}$ describes the probability that a readout binary string $j$ is corrupted into another string $i$ upon measuring. The corruption matrix can be related to the POVMs by $(E_j)_{kl}=\delta_{kl} C_{jl}$. This results in diagonal POVM elements, which clearly is not the most general formulation. However, from a modeling perspective, it is less prone to overfitting on noisy data~\cite{Wold2024}. In addition, measurement errors of real quantum hardware have diminishing off-diagonal elements and are thus well described by the corruption matrix~\cite{Oszmaniec2020}.

To retrieve quantum maps from experiments, we estimate Pauli-string probabilities. In the model, they are predicted to be
\begin{equation}\label{eq:prob}
\hat{p}_j^{\bf{s,b}}(\boldsymbol{\theta})= \sum_{k=1}^{d} C_{jk} \cdot ( P_{\bf{b}}\mathcal{T}(P_{\bf{s}}\rho_{\text{\tiny{0}}} P^{\dag}_{\bf {s}};\boldsymbol{\theta})  P^{\dag}_{\bf{b}})_{kk},
\end{equation}
where the initial state $\rho_0$ and the correction matrix $C$ constitute the SPAM model. First, a Pauli string $P_{\boldsymbol{s}}$ adapts the initial state, followed by the action of the map $\mathcal{T}$. A final Pauli string $P_{\boldsymbol{b}}$ is applied to adapt the measurement basis before probabilities are computed. 
In the experiment, the probabilities~(\ref{eq:prob}) must be estimated by repeated execution and measurements (shots) as they are not directly accessible. This yields a set of estimated probabilities $f^{(\bf{s},\bf{b})}_j$ with sampling error that goes as $\mathcal{O}(\frac{1}{\sqrt{N_s}})$, where $N_s$ is the number of shots. For a map on $n$ qubits, there are $6^n \times 3^n = 18^n$ total Pauli strings. Since the complete set is experimentally prohibitive for even a moderate number of qubits (and practically not necessary for achieving high accuracy), we utilize a random subset of Pauli strings.
By minimizing the mean-square loss function over probabilities
\begin{equation}\label{eq:loss}
    L(\boldsymbol{\theta}) = \sum_{\bf{s},\bf{b},j} 
\left[ \hat{p}^{(\bf{s},\bf{b})}_j(\boldsymbol{\theta}) - f^{(\bf{s},\bf{b})}_j \right]^2,
\end{equation}
with respect to the model parameters $\boldsymbol{\theta}$ by means of a gradient-based method, we obtain a map consistent with the data. Across all experiments, we optimize each map using $4000$ iterations and a learning rate of $10^{-3}.$

\subsection{Bootstrap Error Estimation}
\label{app:BootstrapErrorEstimation}

We employ a bootstrap resampling procedure for estimating the combined error of the map retrieval protocol and finite sampling on hardware. In particular, it is important to establish how the errors propagate to statistics related to the CSRs, as this is a metric particularly sensitive to perturbations.

As described in Sec.~\ref{app:Experimental circuit optimization}, the frequencies $f^{(\bf{s},\bf{b})}_j$ are estimated for a given circuit on hardware using 2000 shots across 6 independent realizations, totaling 12000 shots for each Pauli mode $(\bf{s},\bf{b})$. To emulate multiple experiments with different random finite sampling, we take the experimentally determined $f^{(\bf{s},\bf{b})}_j$ as a ground-truth distribution and resample a new dataset $\tilde{f}^{(\bf{s},\bf{b})}_{j,B}$ with the same number of shots. For each circuit, we generate a total of $N_B = 10$ such datasets and fit independent randomly initialized maps to them. This yields an ensemble of maps $D_{\mathcal{T}} = \{\mathcal{\Tilde{T}}_r^{(i)}\}_{(i=1, \cdots, N_B)}$ that vary due to different finite sampling of data and optimization inaccuracies. For any quantity $F(\mathcal{T})$ that is computed on a map $\mathcal{T}$, we can estimate the ensemble variance as 
\begin{equation}\label{eq:bootstrapVar}
    s_F^2 =\frac{1}{N_B - 1}\sum_{i=1}^{N_B}{[F(\Tilde{\mathcal{T}}}_r^{(i)}) - \Bar{F}(D_{\mathcal{T}})]^2,
\end{equation}
where $\Bar{F}(D_{\mathcal{T}})$ is the average over the ensemble.

We compute averages of quantities across $N_E$ independent experiments as $\Bar{F} = \frac{1}{N_E}\sum_{j=1}^{N_E} F(\mathcal{T}^{(j)})$. To estimate the error, we average and normalize the bootstrap ensemble of each experiment $D^{(j)}_{\mathcal{T}}$ using $10$ resamplings as
\begin{equation}
    \text{err}[\Bar{F}] = \frac{1}{N_E}\sqrt{\sum_{j=1}^{N_E} \text{Var}[F(\mathcal{T}^{(j)})]} \approx  
    \frac{1}{N_E}\sqrt{\sum_{j=1}^{N_E} s_F^{2(j)}},
\end{equation}
where the true variance of the quantity was replaced with its ensemble variance~(\ref{eq:bootstrapVar}). In Sec.~\ref{app:synthetic_benchmarks}, we investigate on synthetic data how well this approximation holds for quantities such as marginal densities of the CSR.

\subsection{Synthetic benchmarks}
\label{app:synthetic_benchmarks}

To benchmark the accuracy of the tomography method, we generate synthetic data sampled from known SPAM error models and quantum maps. We then tomographically retrieve the maps from this data and compare them to the known original maps, on the level of their CSR distributions. We mimic real experiments by generating a synthetic SPAM error model as a convex combination of an ideal and a random SPAM error model. The initial state is defined as 
\begin{equation}\label{eq:synthInit}
    \rho_0 = (1-p_1)\ket{\boldsymbol{0}}\bra{\boldsymbol{0}} + p_1\delta\rho
\end{equation}
and the POVMs as
\begin{equation}\label{eq:synthPOVM}
    E_j = (1-p_2)\ket{\boldsymbol{j}}\bra{\boldsymbol{j}} + p_2\delta E_j.
\end{equation}
In the above expression, $p_1, p_2 \in [0,1]$, and $\delta\rho$ and $\delta E_j$ are sampled randomly in accordance with Eqs.~ (\ref{eq:init}) and (\ref{eq:CM}). We implement target maps $\mathcal{T}(\rho)$ by simulating the integrable and chaotic circuits described in Fig.~\ref{fig:circuit} of the main text for $T=10$. Using the synthetic SPAM errors and target maps, we compute probabilities 
\begin{equation}\label{eq:prob_map}
p_j^{(\bf{s,b})}(\boldsymbol{\theta})= \text{Tr}[E_j P_{\bf{b}}\mathcal{T}(P_{\bf{s}}\rho_{\text{\tiny{0}}} P^{\dag}_{\bf {s}};\boldsymbol{\theta})  P^{\dag}_{\bf{b}}].
\end{equation}
Likewise, for the SPAM model retrieval, we produce probabilities
\begin{equation}\label{eq:prob_spam}
p_j^{(\bf{s})}(\boldsymbol{\theta})= \text{Tr}[E_j P_{\bf{s}}\rho_{\text{\tiny{0}}} P^{\dag}_{\bf {s}}].
\end{equation}

\begin{figure}[t]
     \centering
     \includegraphics[width=\columnwidth]{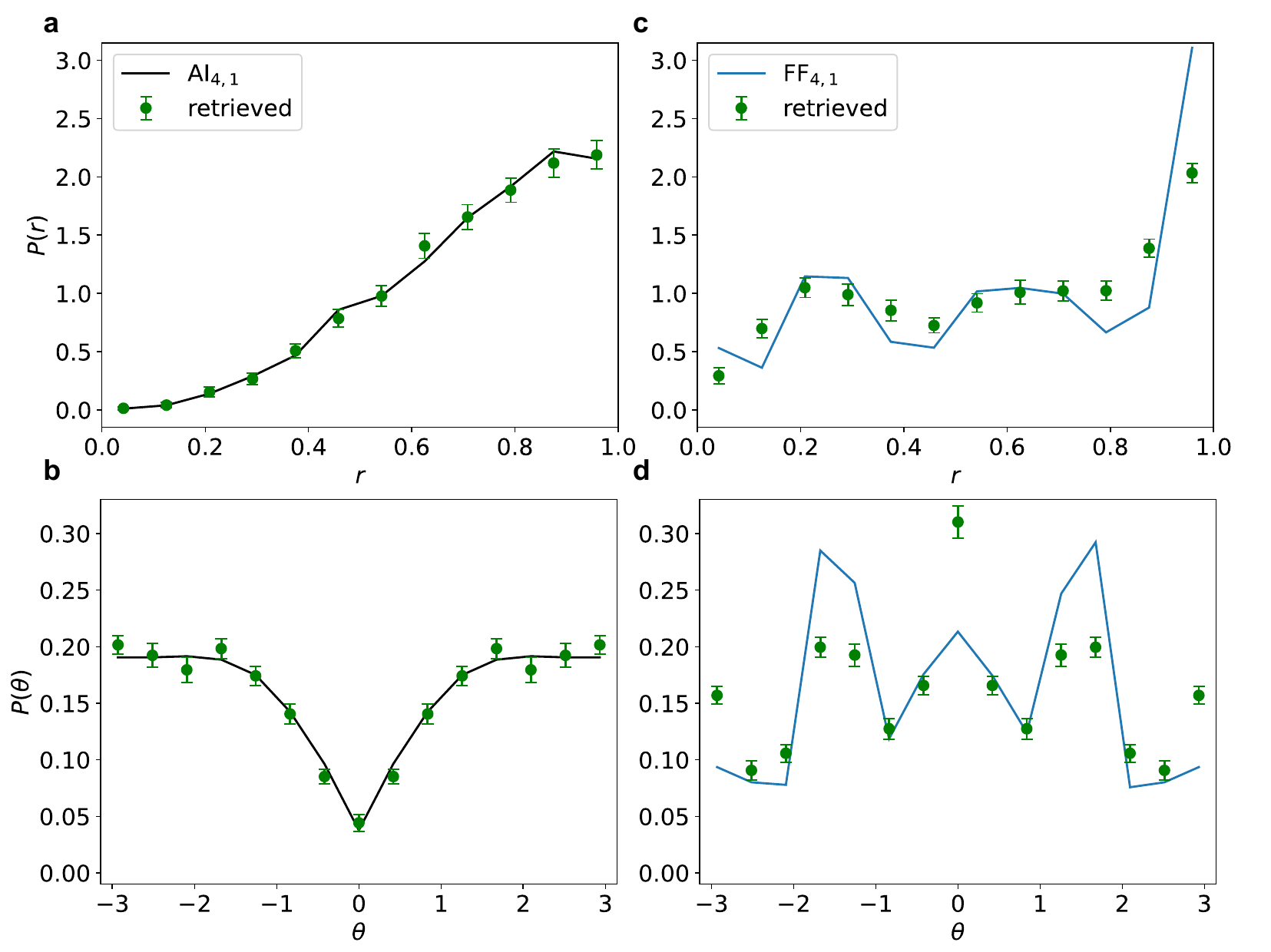}
     \caption{ {\bf Marginal CSR distributions of ideal circuits retrieved from synthetic noisy Pauli-string measurements.} 
     The synthetic data was generated by simulating ten independent realizations of the chaotic ($\textbf{\textsf{a}}$ and $\textbf{\textsf{c}}$) and integrable ($\textbf{\textsf{b}}$ and $\textbf{\textsf{d}}$) circuits, using Eq.~(\ref{eq:prob_map}). $N_s = 12000$ shots and $N_p = 3704$ Pauli modes were used to simulate experimental conditions. The maps were then retrieved using the protocol described in  Sec.~\ref{app:MapRetrival}. For each circuit, the synthetic data was bootstrap resampled and refitted ten times, on which the protocol error was estimated. 
     {\bf\textsf{a}}--{\bf\textsf{b}}, Comparison of the ({\bf\textsf{a}}) marginal radial, $P(r)$, and ({\bf\textsf{b}}) marginal angular, $P(\theta$), distributions retrieved from synthetic data for the chaotic circuit with the ideal statistics of $\text{AI}_{4,1}$.
     {\bf\textsf{c}}--{\bf\textsf{d}}, Comparison of the ({\bf\textsf{c}}) marginal radial, $P(r)$, and ({\bf\textsf{d}}) marginal angular, $P(\theta$), distributions retrieved from synthetic data for the integrable circuit with the ideal statistics of $\text{FF}_{4,1}$.
     }
     \label{fig:synthetic_benchmark}
\end{figure}

In our synthetic benchmark, we use $p_1 = p_2 = 0.05$ for the SPAM error model, overshooting the error rate of the real quantum hardware by several percent. To match the experimental setup, we compute probabilities on a random subsample of $N_p=3704$ Pauli modes $(\bf{s,b})$ for the map, and $N_p = 1296$ Pauli modes $(\bf{s})$ for the SPAM error model. From these we sample and average $N_s = 12000$ shots yielding frequencies $f_j^{(\bf{s,b})}$ and $f_j^{(\bf{s})}$ with finite sampling errors. Finally, we retrieve the SPAM error model and map from $f_j^{(\bf{s})}$ and $f_j^{(\bf{s,b})}$, as described in Sec.~\ref{app:MapRetrival}.
We repeat this procedure for 10 independent realizations of the integrable and chaotic circuits.
The marginal CSR distributions of the retrieved maps are contrasted to the synthetic ground truth in Fig.~\ref{fig:synthetic_benchmark}. The errors are computed as described in Sec.~\ref{app:BootstrapErrorEstimation}.

Figure~\ref{fig:synthetic_benchmark}a--b show that the marginal distributions are all within error bars when retrieving chaotic dynamics, suggesting that no noticeable bias is detected. Furthermore, Fig.~\ref{fig:synthetic_benchmark}c--d show that the integrable marginal distributions of CSR are mostly consistent within error bars between the retrieved and the synthetic ground truth. Densities tend to be underestimated around $r=1$ and $\theta = \pm \frac{\pi}{2}$. Also some overestimation happens around $\theta = \pm \pi$ and $\theta = 0$. This suggests that the retrieval protocol is slightly biased at the level of CSR when retrieving strongly integrable dynamics. 

Calculating the mean of these marginal distributions, we find, for the chaotic circuit, $\langle r \rangle = 0.725 \pm 0.003$ (ideal 0.727) and $\langle -\cos{\theta} \rangle = 0.18 \pm 0.01$ (ideal 0.17), consistent with no detectable bias. For the integrable circuit, $\langle r \rangle = 0.592 \pm 0.004$ (ideal $0.616$) and $\langle -\cos{\theta} \rangle = -0.126 \pm 0.008$ (ideal $-0.171$), confirming a small bias for the integrable case.

Even though some bias was confirmed in the integrable case, these benchmarks show that the protocol is not a limiting factor for accurately retrieving and discriminating between integrable and chaotic dynamics, even when data is limited.

\section{Theoretical details}

\subsection{Numerical implementation of dissipative quantum circuits}
\label{app:numerical_implementation}

Let $U$ be a unitary evolution operator on $L$ qubits. We divide the $L$ qubits into $n$ system qubits, which are not traced out (i.e., erased by the environment), and $e$ environment (or ancilla) qubits, which are. If we initialize all the environment qubits in $|0\rangle$, then we have $2^e$ Kraus operators $K_{j_1,\dots,j_e}=\langle j_1\cdots j_e|U|0_1\cdots0_e\rangle$ ($j_1,\dots,j_e\in\{0,1\}$). In a basis where all system qubits come after the environment qubits, this corresponds to partitioning the unitary matrix $U$ into a $2^e\times 2^e$ block matrix, with each block $2^n\times 2^n$. The blocks in the first column are the Kraus operators.
The dissipative quantum circuit is then represented by the $4^n\times4^n$ quantum channel
\begin{equation}
\label{eq:def_Kraus}
    \Lambda=\sum_{j_1,\dots,j_e=0}^1 K_{j_1,\dots,j_e}\otimes K_{j_1,\dots,j_e}^*.
\end{equation}

In the main text, we implemented $U$ as either a (regular) FF circuit (Fig.~\ref{fig:circuit}a) or a chaotic circuit (Fig.~\ref{fig:circuit}b), which are parametrized by $n$, $e$ (such that $n+e=L$), and the depth $T$ of the circuit. Accordingly, we denote the respective ensemble of maps by FF$_{n,e}$ and AI$_{n,e}$, respectively.
As $T\to\infty$, the chaotic circuit approaches a Haar-random unitary of the same dimension. We thus sampled $U$ as a $2^{L}\times 2^{L}$ matrix from the circular unitary ensemble (CUE).
To generate the FF circuit, we proceeded as follows. First, we write $U=e^H$, where
\begin{equation}
    H=\sum_{i>j=1}^{2L}\gamma_i J_{ij} \gamma_j.
\end{equation}
is the corresponding FF Hamiltonian in the Majorana representation. $\gamma_i=\gamma_i^\dagger$ ($i=1,\dots, 2L$) are gamma matrices, which satisfy the Clifford algebra $\{\gamma_i,\gamma_j\}=\delta_{ij}$ and represent the Majorana fermions. In the computational basis, the gamma matrices are written as ($i=1,\dots,L)$:
\begin{equation}
    \gamma_{2i-1}=\frac{1}{2^{L/2}}\prod_{j=1}^{i-1}\sigma^z_j \sigma_i^x,
    \quad
    \gamma_{2i}=\frac{1}{2^{L/2}}\prod_{j=1}^{i-1}\sigma^z_j \sigma_i^y.
\end{equation}
Fermionic systems without any symmetries (besides fermion parity, which is always present) belong to class D~\cite{Altland1997}. Hence, $J$ is an imaginary antisymmetric matrix without any additional symmetries. We set $J=\log O$ and sample $O$ from the Haar measure on SO($2L$). Generating $O$ instead of $J$ guarantees that we sample uniformly over FF circuits, instead of giving a preference to unitaries close to the identity. The FF circuit realized in the experiments has an additional U(1) conservation law, $[H,Q]=0$ with $Q=\sum_{i=1}^L \sigma_j^z$ (i.e., ``particle-number'', see Sec.~\ref{app:FF_Symmetries} for additional details). To implement this symmetry, we sum the projection of $H$ into all sectors of fixed quantum numbers, i.e.,
\begin{equation}
    H\to \sum_{q=0}^L \mathbb{P}_qH\mathbb{P}_q,
\end{equation}
where 
\begin{equation}
    \mathbb{P}_q=\frac{1}{L+1}\sum_{s=0}^L e^{2\pi i s (Q-q)/(L+1)}
\end{equation}
is the projector into a sector of fixed quantum number $q$.

\begin{figure*}[t]
     \centering
     \includegraphics[width=\textwidth]{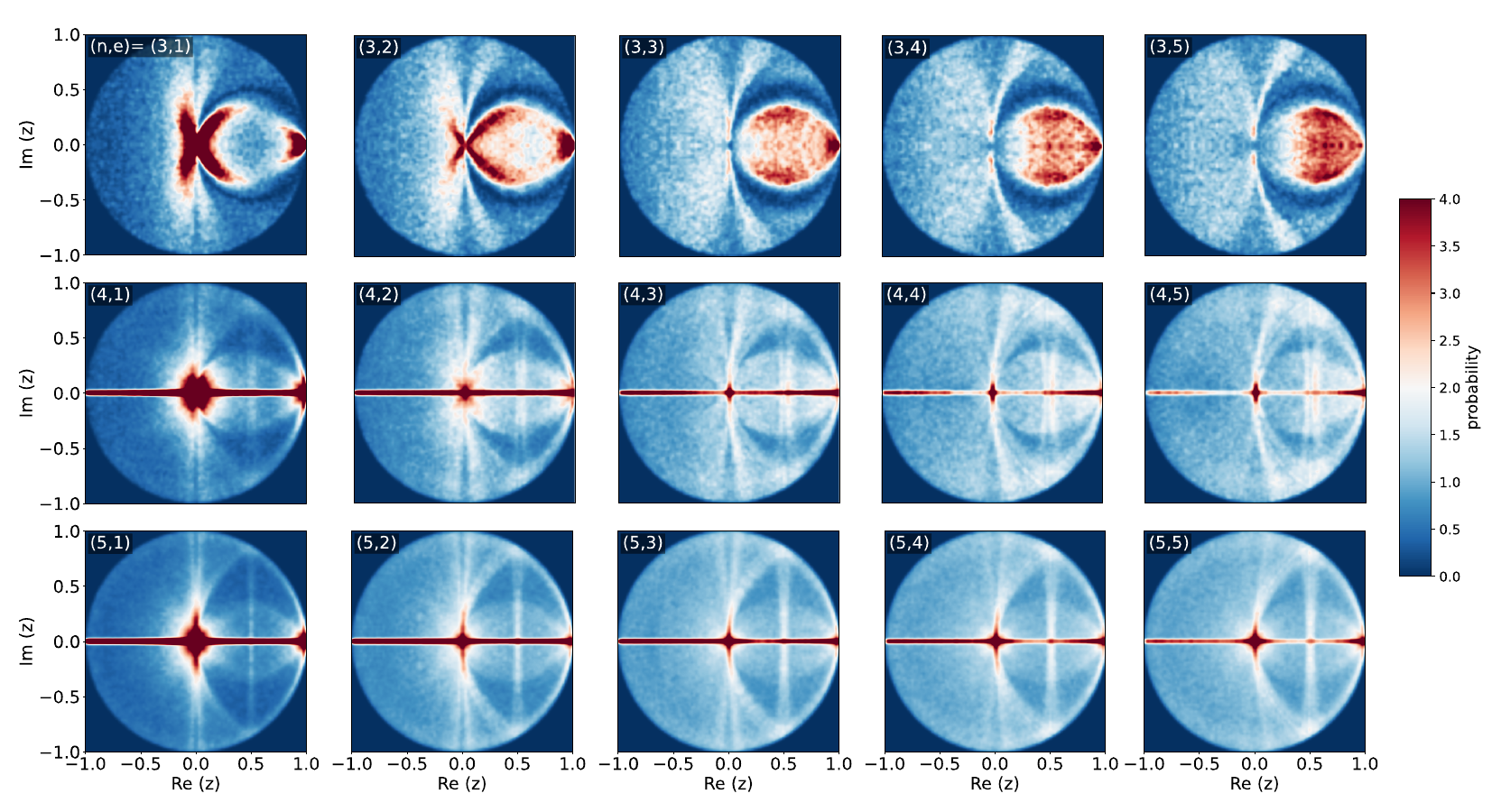}
     \caption{{\bf Numerical CSR distribution $P(z)$ of Haar-random free-fermion channels for different numbers of system qubits $n\in[3,5]$ and environment qubits $e\in[1,5]$.} We consider only the steady-state sector $\Lambda_0$, see Sec.~\ref{app:FF_Symmetries}, averaged over $10^4$ realizations of the channel and discarded eigenvalues within a distance $0.01$ of the real axis.}
     \label{fig:finite_size_FF}
\end{figure*}

\subsection{Finite-size effects in the CSR distribution} 
\label{app:finite-size}

Following the procedure described above, we computed the CSR distribution of $\Lambda$ for FF$_{n,e}$ with different values of $n$ and $e$. The results are shown in Fig.~\ref{fig:finite_size_FF}. We see that, as $n$ and $e$ increase, the distribution approaches the flat Poisson distribution, characteristic of ``generic'' regular spectra. On the other hand, there are pronounced finite-$n$ and finite-$e$ effects. In particular, for $e=1$ as in our experiments, the distribution concentrates around the real axis, with a large peak around the origin. We conclude that the ``candle''-like distribution is a finite-size effect and, if larger experiments were performed, we would converge to Poisson statistics. Nevertheless, these finite-size effects actually help in the experimental detection of dissipative integrability, as they render the distribution even more distinct from the chaotic one.

\subsection{Symmetries in FF circuits}
\label{app:FF_Symmetries}

When performing a level statistics analysis, all symmetries need to be resolved before the statistics (say, CSRs) are computed. That is, if the quantum channel $\Lambda$ has a symmetry $\mathcal{Q}$, $[\Lambda,\mathcal{Q}]=0$, then, in the eigenbasis of $\mathcal{Q}$, $\Lambda$ is block diagonal and the eigenvalues coming from different sectors (i.e., blocks) are independent. Each sector must be analyzed independently. 

The chaotic circuit depicted in Fig.~\ref{fig:circuit}b does not have any unitary symmetry. On the other hand, the FF circuit in Fig.~\ref{fig:circuit}a has a U(1) symmetry (particle-number conservation), which we analyze in the following.
To make the U(1) symmetry of the FF circuit manifest, we note that the gates can be explicitly written as
\begin{equation}
    \sqrt{\mathrm{iSWAP}}=e^{\mathrm{i}\frac{\pi}{8}\left(\sigma^x\otimes\sigma^x+\sigma^y\otimes\sigma^y\right)},
    \qquad Z=e^{\mathrm{i}\frac{\theta}{2}\sigma^z},
\end{equation}
which commute with $\sigma^z\otimes I_1+I_1\otimes \sigma^z$, where $I_k$ is the identity matrix on $k$ qubits. By composition, the full unitary circuit $U$ on $L$ qubits commutes with $Q=\sum_{i=1}^L\sigma_i^z$. 

\begin{figure}[t]
     \centering
     \includegraphics[width=\columnwidth]{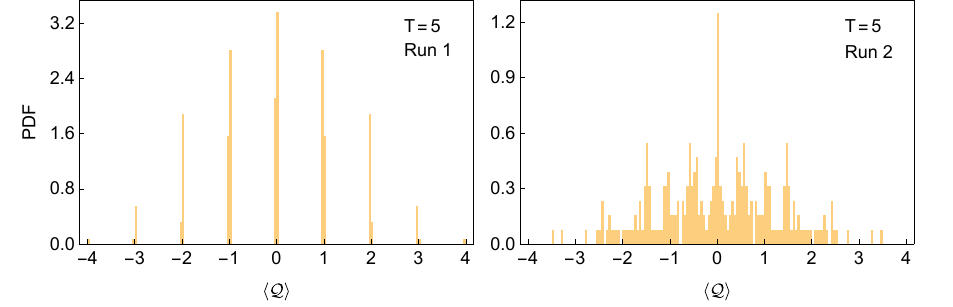}
     \caption{{\bf Distribution of the expectation value $\langle \mathcal{Q} \rangle$ in the eigenstates of the quantum map $\Lambda$ for two different runs with fixed depth $T=5$.} The ideal circuit conserves particle number exactly and the distribution is a sum of delta peaks, Eq.~(\ref{eq:dist_numb_binomial}). Due to the intrinsic noise in the hardware, experimentally, the symmetry is only approximate and the delta peaks broaden. For shallow circuits, the symmetry breaking is usually weak (left panel), but for a few runs, there can be substantial overlap of the sectors (right panel), which we discarded.}
     \label{fig:numberconservation5}
\end{figure}

Because the system and environment qubits can exchange ``particles'' (i.e., logical 0s and 1s), the Kraus operators defined in Eq.~(\ref{eq:def_Kraus}) do not commute with $Q_n=\sum_{j=e+1}^L \sigma^z_j$, where the sum is over the system qubits. Instead, the dissipative quantum circuit has a weak U(1) symmetry~\cite{buca2012} and commutes only with \begin{equation}
    \mathcal{Q}_n=\frac{1}{2}\left(Q_n\otimes I_n-I_n\otimes Q_n\right),
\end{equation}
i.e., only the difference between the number of ``particles'' in the ket and the bra of the system density matrix is conserved. $\mathcal{Q}_n$ has $2n+1$ distinct eigenvalues $q=-n,-n+1,\dots, n$, which label the weak symmetry sectors of $\Lambda$, $\Lambda=\bigoplus_q \Lambda_q$, where $\Lambda_q$ is a block of dimension $\binom{2n}{n+q}$. 

The symmetry sectors can be separated by either preselection or postselection, and we proceeded with the latter. We diagonalize $\Lambda$ as
\begin{equation}
    \Lambda=\sum_{\alpha=1}^{4^n} |R_\alpha\rangle \lambda_\alpha \langle L_\alpha|,
\end{equation}
where $|R_\alpha\rangle$ ($|L_\alpha\rangle$) are the right (left) eigenvectors with eigenvalue $\lambda_\alpha$, $\Lambda|R_\alpha\rangle=\lambda_\alpha |R_\alpha\rangle$ ($\langle L_\alpha|\Lambda=\lambda_\alpha \langle L_\alpha|$). For each eigenvalue $\lambda_\alpha$, we compute $\langle\mathcal{Q}\rangle_\alpha=\langle L_\alpha|\mathcal{Q}_n |R_\alpha\rangle\in\{-n,\dots,n\}$. If $\langle\mathcal{Q}\rangle_\alpha=q$, then $\lambda_\alpha$ is an eigenvalue in $\Lambda_q$.

\begin{figure*}[tbp]
     \centering
     \includegraphics[width=\textwidth]{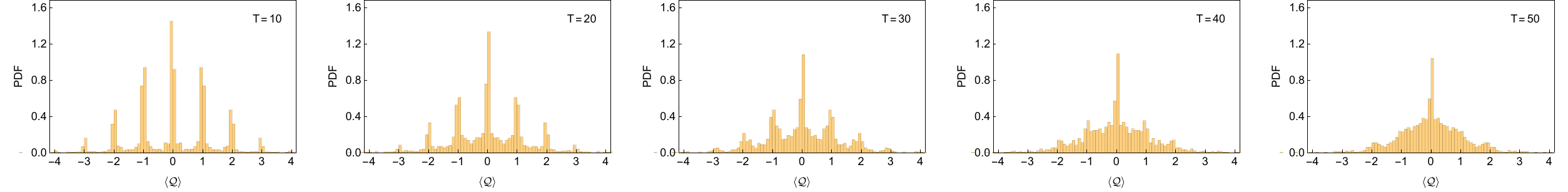}
     \caption{{\bf Distribution of the expectation value $\langle \mathcal{Q} \rangle$ in the eigenstates of the quantum map $\Lambda$ for different circuit depths $T$.} As the depth increases, the symmetry is broken more strongly and different sectors overlap considerably. For any $T\geq10$ the symmetry is sufficiently broken that we do not need to resolve the symmetry.}
     \label{fig:numberdegradation}
\end{figure*}

\subsection{Approximate Symmetries}
\label{app:approximate_symmetries}

While the circuit that is theoretically implemented might have a given symmetry, the intrinsic noise of the hardware will not respect it, thus mixing different sectors. However, if the noise is weak and the circuit is sufficiently shallow, the symmetry is only weakly broken and states in different sectors hybridize little. In that case, we should still consider individual symmetry sectors when performing the spectral analysis; if the sectors mix strongly, all eigenvalues of $\Lambda$ should be considered. To quantify the degree of mixing, we computed the distribution of $\langle\mathcal{Q}\rangle_\alpha$. If the implementation were perfect (without any noise), we would have a binomial distribution
\begin{equation}
\label{eq:dist_numb_binomial}
    P(\langle\mathcal{Q}\rangle)=
    \frac{1}{4^n}\sum_{q=-n}^{n}\binom{2n}{q}\delta_{q,\langle\mathcal{Q}\rangle}.
\end{equation}
When the symmetry is weakly broken, the delta peaks are slightly smeared, see Fig.~\ref{fig:numberconservation5}a, but the sectors can still be clearly separated. When the symmetry is strongly broken, see Fig.~\ref{fig:numberconservation5}b, then the sectors all mix and cannot be separated. Note that, as is the case in Fig.~\ref{fig:numberconservation5}, different parameter runs of the same circuit can break the symmetry differently. Finally, we investigate how the symmetry degrades as we increase the circuit depth. In Fig.~\ref{fig:numberdegradation} we show the distribution of $\langle\mathcal{Q}\rangle$ for each depth $T=10,\dots,50$ (corresponding to the CSR shown in Fig.~\ref{fig:integrable_vs_chaos} of the main text) averaged over the 10 samples that we obtained experimentally. We see that for any $T\geq10$ the sectors are (partially) mixed; hence, in performing the statistical analysis of Fig.~\ref{fig:integrable_vs_chaos}, we did not separate symmetry sectors.

\end{document}